\documentstyle[preprint,prb,aps,amssymb]{revtex}

\newcommand{\bmq}{{\mbox{\boldmath $q$}}}

\begin{document}
\preprint{WIS-99/24 June DPP}
\draft

\date{\today}
\title{Applications of computed Nuclear Structure Functions to Inclusive
Scattering, $R$-ratios and their Moments}
\author{A.S. Rinat}
\address{Department of Particle Physics, Weizmann Institute of Science,
         Rehovot 7616, Israel\footnote{To be published in Proceedings
         of $'$Prospects of Hadron and Nuclear Physics$'$, May 1999,
         Trieste, Italy.}}

\maketitle
\begin{abstract}

We discuss applications of previously computed  nuclear structure functions (SF)
to inclusive cross  sections, compare predictions with recent CEBAF
data and perform two scaling tests. We mention that
the large $Q^2$ plateau of scaling functions may only in part
be due to the asymptotic limit of SF, which prevents the
extraction of the nucleon momentum distribution in a model-independent way.
We show that there may be sizable discrepancies between computed
and semi-heuristic estimates of SF ratios. We compute
ratios of moments of nuclear SF and
show these to be in reasonable agreement with data. We speculate that an
effective theory may underly the model for the nuclear SF, which produces
overall agreement with several observables.

\end{abstract}

\newpage

\section{Inclusive cross sections.}

In the following we discuss three applications of previously computed
nuclear structure functions  $F_k^A$:

i) Cross sections for inclusive scattering of high-energy leptons from nuclei
and associated scaling tests.

ii) Comparison of computed $R$ ratios and previously used, semi-heuristic
methods aimed  to isolate  $F_2^A$

iii) Determination of moments of the above nuclear structure functions (SF)
and comparison of their ratios with data.

We start with the cross  section per nucleon of a one-photon induced process
\begin{eqnarray}
\frac{d^2\sigma_{eA}(E;\theta,\nu)/A}{d\Omega\,d\nu}
=\frac{2}{M}\sigma_M(E;\theta,\nu)
\bigg\lbrack\frac {xM^2}{Q^2}F_2^A(x,Q^2)+
{\rm tan}^2(\theta/2)F_1^A(x,Q^2)\bigg\rbrack
\label{a1}
\end{eqnarray}
The inclusive, as  well as the Mott cross  for point-nucleons $\sigma_M$,
appear above as functions of the  beam energy $E$, the  scattering angle
$\theta$ and the energy loss $\nu$,
but may also be expressed in alternative kinematic variables.

The two SF $F_k^A$ above describe the scattering  of unpolarized
electrons from  randomly oriented targets and have been expressed
as functions of the squared 4-momentum $Q^2=\bmq^2-\nu^2$
and the Bjorken variable  $x=Q^2/2M\nu$. We shall
analyze recent data on inclusive scattering of 4.05 GeV electrons
on various targets \cite{arr}. The ranges of scattering angles $15\lesssim
\theta\lesssim 74$ and the measured energy losses $\nu$
correspond to $1\lesssim Q^2({\rm  GeV}^2)\lesssim 7$  and
$0.20\lesssim  x  \lesssim 4.2$,  vastly  extending  the kinematic limits
of  the  older   NE3  SLAC $\,$ \cite{day}  and  related
experiments \cite{brad,arr1}.  On the  deep-inelastic side
$0.2\lesssim x\lesssim 1$ those overlap with the $'$classical$'$ EMC domain.
Certainly there a description ought to include the quark-gluon
sub-structure of the nucleon.

In the past  we have proposed and applied
a model, which relates SF's $F_k^A$ and $F_k^N$
of  a nucleus and a nucleon, by means of a SF $f^{PN}$ of a
nucleus composed of point-nucleons.
Disregarding virtual pions etc., one has  \cite{gr}
\begin{eqnarray}
F_k^A(x,Q^2)=\int_x^A  \frac{dz}{z^{2-k}}f^{PN}(z,Q^2)
F_k^N \bigg (\frac{x}{z},Q^2\bigg   )
\label{a2}
\end{eqnarray}

A similar equation for  momentum fractions has been proposed before
\cite{aku}. Those quantities tend in the $Q^2\to\infty$ limit to Bjorken
variables and (\ref{a2})  states the approximate validity  for large, finite
$Q^2$: Its quality will deteriorate with decreasing $Q^2$.

A calculation of  the above nuclear SF $F_k^A$ rests on two input elements. The
first is the averaged SF of a nucleon $F_k^N\equiv F_k^{<p,n>}$,
properly weighted with proton and neutron fractions in the nucleus.
The non-perturbative model leading to (\ref{a2}) prescribes
the $N$ to be on-mass shell, and its SF's are therefore known \cite{bod}.

The second element in (\ref{a2}) is the SF for
a nucleus, composed of point-nucleons $f^{PN}$ and which
accounts for  nuclear  dynamics.
It is calculable \cite{rt1,rt2} in a relativistic extension of the
non-relativistic
(NR) Gersch-Rodriguez-Smith  (GRS) series in $1/q\,\,\,$  \cite{grs}.
The  latter   contains  an   asymptotic  limit   (AL),  related   to  the
single-nucleon momentum distribution (MD) $n(p)$, and Final State Interactions
(FSI), dominated  by hard binary  collisions between  the knocked-on
nucleon and a nucleon from the core.

We remark that the model leading to (\ref{a2}) locates weak
$A$-dependence
of  $F_k^A(x,Q^2)$  in  the  neutron excess  $\delta  N/2A$, and  in
$f^{PN}\,\,$ \cite{rt1}, thus
\begin{eqnarray}
F_k^A(x,Q^2)\approx F_k(x,Q^2) +{\cal O}(1/A);\,\,\,A \gtrsim 12,
\label{a3}
\end{eqnarray}

The above mentioned input allows predictions to be made for the cross sections
(\ref{a1}) \cite{rt1,rt2}. From a comparison  with the new CEBAF
data on Fe  $\,$ \cite{arr} in Fig. 1 one concludes:

i) For all but the smallest $Q^2$, there is good agreement  in the
(deep-)inelastic region $\nu>Q^2/2M,\,x<1$
and  satisfactory  correspondence  on  the  nucleon elastic (NE)
side  $x\gtrsim 1$, contiguous to the quasi-elastic peak (QEP).

ii) Since $Q^2$  increases with  $\theta$, and  for given  $\theta$ with
decreasing $\nu$, one observes  that  discrepancies grow
with decreasing $\theta$, i.e. for decreasing $Q^2$, as expected. One estimates
$Q^2_c(x,\theta)\approx 1.5\,\,{\rm GeV}^2$,   below    which   the
representation (\ref{a2}) may become progressively flawed.

iii) For each $\theta$,  cross sections for the lowest energy losses $\nu$
drop orders of magnitude from their maximum. Theory overestimates the data
there by a factor up to 2-3.

In spite of the above, it is not  at all  clear that in the latter
regions there is a real discrepancy: Alternative MD $n(p)$ for Fe,
produce results which range over the area of the above mentioned local
discrepancies without spoiling the agreement for higher $\nu$ (see
Figs. 5,6 in Ref. \onlinecite{rt1}).

The NE3 experiment \cite{day} has also been analysed by means of versions
of the  Plane Wave Impulse  Approximation (PWIA)  in terms of  a spectral
function,     occasionally     supplemented     by     additional     FSI
\cite{om,ciof1,ciof2,oset}, e.g. $2p-1h$  FSI on the PWIA \cite{ciof3}.
The  above mentioned  GRS and IA approaches agree very well with data, except
for the smallest $\nu$, where
Ciofi and Simula somewhat underestimate intensities \cite{ciof3}, while
our approach overestimates those. The otherwise surprising correspondence
can be
understood in the light of a recent proof, that a NR version of these two,
quite different theories agree order-by-order in $1/q\,$ \cite{rj}.

We turn to scaling analyses, previously applied to the NE3 data
$\,$\cite{arr1,rt1}.
We first consider ratios of inclusive cross sections for different targets
under identical kinematic conditions
\begin{eqnarray}
\xi^{A_1,A_2}=\bigg (\frac{d^2\sigma^{eA_1}}{A_1}\bigg ) \bigg /
\bigg (\frac {d^2\sigma^{eA_2}}{A_2}\bigg ),
\label{a5}
\end{eqnarray}
for instance using  a relativistic GRS-West scaling variable
suggested by Gurvitz \cite{sag} or the related $x$
\begin{eqnarray}
y_G\approx (M/q)(\nu-Q^2/2M)\approx (Mq/\nu)(1-x)
\label{a4}
\end{eqnarray}
Originally the analysis had been limited to the region
$y_G<0$ below the QEP, and universal scaling for all $A\gtrsim 12$
had been observed.  However, Eq. (\ref{a3}) holds for all, kinematically allowed
$y_G$, including the (deep-)inelastic region $y_G>0$. In spite of
4-5 orders of magnitude variations of cross sections, one has
for $all\,\,y_G$, and independent of $E,x,Q^2$,
$\xi^{A_1,A_2}(E,x,Q^2)\approx 1$ within 15-20\%, and
frequently  better. Occasional larger deviations for  data with
lowest intensity are readily ascribed to experimental uncertainties.
Table  I illustrates the above for the pairs C,Fe and Fe,Au.

Next we focus on the EMC ratio $\xi^{A,N}_{NE}$, with
$A_2\to\langle N\rangle\approx D/2$. Both the NE region, where the nucleon
remains intact, and the nucleon inelastic (NI) region, describing
excitation or fragmentation of the $N$, contain
information on the single-nucleon MD $n(p)$, implicit
in $f^{PN}$, Eq. (\ref{a2}). However, simple  expressions for $\xi^{A,N}$
can only be given for $y_G<0$. Densities, MD
and pair-distribution functions $g_2$ are different
for the $D$ and $A  \gtrsim  12$ and consequently, $\xi^{A,N}_{NE}$
is  not a special case  of  $\xi^{A_1,A_2}$.

Unfortunately the true NE  part does not coincide with $y_G\le 0$ which
one wishes to investigate. As  had already  been realized  in the
analysis of early high-$E$ experiments on  the lightest nuclei D, $^3$He,
$^4$He$\,$ \cite{day1}, even on the elastic  side $y_G<0$  ($x>1$)
of the  QEP (cf. Ref.  \onlinecite{rt2}, Fig. 4), NI and
NE  contributions compete.
Consequently the NE part is not directly
observable and  a scaling analysis of $\xi^{A,N}_{NE}$ requires
NI parts to be removed from  the data. For instance, Eq. (\ref{a2})
provides a model for NI parts, which appear to be in perfect agreement
with the data for $y_G>0$: we assume the same for $y_G\l 0$.

The  procedure  becomes impractical for $y_G\lesssim -0.25$,
where both  total and  NI parts  decrease rapidly.
There   one  has  to  rely on  directly
calculated  NE parts,  using  again  (\ref{a2}), now  with  NE parts  for
$F_k^N$.  The result in terms of static form factors reads
\begin{mathletters}
\label{a6}
\begin{eqnarray}
F_1^{N(NE)}(x,Q^2)&=&\frac{x}{2}[G^N_M(Q^2)]^2\delta(x-1)
\nonumber\\
F_2^{N(NE)}(x,Q^2)&=&\frac{[G^N_E(Q^2)]^2+\eta[G_M^N(Q^2)]^2}{1+\eta}
\delta(x-1)
\label{a6a}\\
F_1^{A(NE)}(x,Q^2)&=&\frac{1}{2}f^{PN}(x,Q^2)[G^N_M(Q^2)]^2
\nonumber\\
F_2^{A(NE)}(x,Q^2)&=&xf^{PN}(x,Q^2)
\frac{[G^N_E(Q^2)]^2+\eta[G_M^N(Q^2)]^2}{1+\eta}
\label{a6b}
\end{eqnarray}
\end{mathletters}
Substitution in (1) yields expressions for the $NE$ parts of cross
sections and consequently for the corresponding parts of
$\xi^{A,N}_{NE}$ ($\eta=Q^2/4M^2$)
\begin{mathletters}
\label{a7}
\begin{eqnarray}
\xi^{A,N}_{NE}(x,Q^2)=f^{PN}(x,Q^2)
&&\bigg \lbrack \frac
{(x^2m^2/Q^2)\frac{[G^N_E]^2+\eta [G_M^N]^2}{(1+\eta}
+{\rm tan}^2(\theta/2) ([G_M^N]^2}
{(m^2/Q^2)\frac{[G^N_E]^2+\eta[G_M^N]^2}{1+\eta}
+{\rm tan}^2(\theta/2)[G_M^N]^2}\bigg \rbrack
\label{a7a}\\
\xi^{A,N}_{NE}(x\approx 1,Q^2)&=&f^{PN}(x\approx 1,Q^2),
\label{a7b}
\end{eqnarray}
\end{mathletters}
where arguments on $G^N$ have been dropped.

Fig. 2 displays  $\xi_{NE}^{Fe,N}(y_G<0,Q^2)$ against  $Q^2$ for  a
number  of narrowly  binned $y_G$ data.   Whenever possible,  we give  the resul
for the described procedures which, with  the exception of $y_G=0$,
approximately agree.  For
all  $y_G<0$  in the  kinematic  range of  the experiment,  $\xi_{NE}$
approaches  the  AL  $Q^2\to\infty$  from  above,  or  can confidently be
extrapolated  to $Q^2$  beyond the  observed ones.   For the
largest $|y_G| $, there is hardly any $Q^2$ dependence.

An observed plateau is  conventionally related to the AL, from which one wishes
to extract  the MD $n(p)$.  However,
the standard argument  becomes invalid, if parts of the  FSI happen to be
weakly $Q^2$-dependent,  causing the  plateau to  also contain FSI parts.
There are strong indications that this might be the case for
the kinematic  range of the  CEBAF experiment .
The interest in scaling analyses may well wane, if the AL cannot  be
separated from $Q^2$-independent FSI, thereby blocking a model-independent
extraction of $n(p)$ (see Ref. \onlinecite{rt4} for details and
an explanation for the seeming contradiction in the
behaviour of $\xi^{A,N}_{NE}$  for low $Q^2$ as seen  in Fig. 2).

\section{R ratios.}

Our second topic deals with the  separation of the two nuclear SF $F_k^A$ in
(\ref{a1}) and more specifically, with the isolation of the dominant $F_2^A$.
This requires data for fixed $x,Q^2$ at different scattering angles
$\theta$ or beam energies $E$, and those are not frequently  available.
Approximate methods start with alternative expressions for the cross
section ratios
\begin{mathletters}
\label{a8}
\begin{eqnarray}
\frac{d^2\sigma_{eA}(E;\theta,\nu)/[A\sigma_M(E;\theta,\nu)]}{d\Omega\,d\nu}
&=&\frac{2Mx}{Q^2}F_2^A(x,Q^2)\bigg \lbrack
1+\frac{2\bigg(1+Q^2/4M^2x^2\bigg)
{\rm tan}^2(\theta/2)}{1+R(x,Q^2)} \bigg \rbrack
\label{a8a}\\
&=&\frac{2Mx}{Q^2}F_2^A(x,Q^2)\bigg\lbrack
1+\frac{Q^2}{2M^2x^2}\kappa(x,Q^2){\rm tan}^2(\theta/2)\bigg \rbrack,
\label{a8b}
\end{eqnarray}
\end{mathletters}
where
\begin{mathletters}
\label{a9}
\begin{eqnarray}
R^A=d^2\sigma_L/d^2\sigma_T&=&\bigg (1+\frac{4M^2x^2}{Q^2}\bigg )\frac{1}
{\kappa^A(x,Q^2)}-1
\label{a9a}\\
\kappa^A(x,Q^2) &=&\frac{2xF^A_1(x,Q^2)}{F^A_2(x,Q^2)}
\label{a9b}
\end{eqnarray}
\end{mathletters}
$R$ is  the ratio of cross sections for the scattering of longitudinal and
transverse photons. It is related to the ratio of the two SF in
$\kappa^A(x,Q^2)$, Eq. (\ref{a13}), which we call the nuclear Callen-Gross (CG)
function. Its dependence on $A\gtrsim 12$ follows from (\ref{a3}) and reads
\begin{eqnarray}
\kappa^A=\kappa^{\langle N\rangle} +{\cal O}(1/A)&\approx&
\kappa^D(x,Q^2)+{\cal O}(1/A)
\nonumber\\
R^A(x,Q^2)&\approx& R(x,Q^2)+{\cal O}(1/A),
\label{a10}
\end{eqnarray}
which agrees with data \cite{arneo,das1}. Recalling the CG $relation$ for
nucleons
\begin{eqnarray}
\epsilon^N_{CG}=\lim_{Q^2\to \infty} \kappa^N(x,Q^2)=1,
\label{a11}
\end{eqnarray}
one finds from (\ref{a13}) and (\ref{a14}) its nuclear analog
\begin{eqnarray}
\epsilon^A_{CG}=\lim_{Q^2\to \infty} \kappa^A(x,Q^2)=1+{\cal O}(1/A)
\label{a12}
\end{eqnarray}
The latter relation can also be proven directly from (\ref{a2}),
using (\ref{13}), whereas the
equality of
nuclear and nucleonic  CG $functions$  (\ref{a15}) and (\ref{a16})
is compatible  with (\ref{a2}),
but does not follow from it.

First we mention an observation for the computed CG functions in the range
(0.2-0.3) $\lesssim x \lesssim (0.7-0.75);\,\,Q^2\ge 5 {\rm GeV}^2$
\begin{eqnarray}
|\kappa^A(x,Q^2)-1|\approx(0.11-0.12),
\label{a13}
\end{eqnarray}
i.e. CG functions in those ranges are close to their asymptotic limit, the
nuclear CG relation (\ref{a12}). Without apparent  cause,
theory predicts a  sign change in $\kappa-1$
at $x_s\approx 0.5-0.6$, which is in
agreement with  data from high energy  $\nu,\bar\nu$ inclusive scattering
(see Fig.  18 in  Ref. \onlinecite{berg}).

The following  remarks relate to computed CG functions (\ref{a13}):

i) Disregard of other than valence quarks requires smoothing of $F_k^N$
for $x\lesssim$ 0.15-0.20,  which entails the same for  $F_k^A$.

ii) Eq. (\ref{a2})  shows that beyond $x\approx 1$,
$f^{PN}$ draws on an  ever smaller support of dwindling intensity and  accuracy,
rendering unreliable $F_k^A(x,Q^2)$, and thus $\kappa(x,Q^2)$, for
$x\gtrsim 1.3$.

We briefly mention  approximations  $R_n$ for $R^A\approx R$, or
for the CG function  $\kappa_n$.  For those one has from (\ref{a9})
\begin{eqnarray}
R(x,Q^2)=\beta_n(x,Q^2)R_n(x,Q^2)+\bigg (\beta_n(x,Q^2)-1   \bigg)
\label{a14}
\end{eqnarray}
Deviations of $\beta_n(x,Q^2)=\kappa_n(x,Q^2)/\kappa(x,Q^2)$ from
1  determine the quality of the approximation.

A) High-$Q^2$ approximation for $1 \lesssim x\lesssim $0.6: $\kappa_L=
[\beta_L]^{-1}=1$,
\begin{mathletters}
\label{a15}
\begin{eqnarray}
R^{comp}(x,Q^2)&=&
\beta_L(x,Q^2)R_L(x,Q^2)+\bigg (\beta_L(x,Q^2)-1 \bigg)
\label{a15a}\\
&\approx& R_L(x,Q^2)+\bigg (\beta_L(x,Q^2)-1\bigg )
\label{a15b}\\
R_L^{(1)}(x,Q^2)&=&\frac{4M^2x^2}{Q^2}+ (\beta_L(x,Q^2)-1)
\label{a15c}\\
R_L^{(2)}(x,Q^2)&=&\frac{4M^2x^2}{Q^2},
\label{a15d}
\end{eqnarray}
\end{mathletters}
with (\ref{a19}) the result, computed from Eqs.  (\ref{a2}).

B) NE approximation  for $x\approx1$: Using (\ref{a7}) and exploiting
in (\ref{a6}) the approximate scaling of static electro-magnetic form factors,
one has $1/[(\mu_M^p)^2+(\mu_M^n)^2]=0.0874\,$ \cite{bos} and thus
\begin{eqnarray}
\kappa^A_{NE}&=&2xF_1^{A(NE)}/F_2^{A(NE)}
\nonumber\\
&\approx&(0.0874+\eta)/(1+\eta)
\label{a16}
\end{eqnarray}
Inserting (\ref{a23}) into (\ref{a18}) gives ($Q^2$ in GeV$^2$)
\begin{mathletters}
\label{a17}
\begin{eqnarray}
R(x,Q^2)&=&\beta_{NE}(x,Q^2)R_{NE}(x,Q^2)
+\bigg (\beta_{NE}(x,Q^2)-1 \bigg)
\label{a17a}\\
R_{NE}^{(1)}(x,Q^2)
&=& \frac{0.31}{Q^2}+\bigg (\frac{0.31}{Q^2}+1\bigg )
\bigg ( \frac{x^2-1}{1+\eta}\bigg ),
\label{a17b}\\
R^{(2)}_{NE}(x,Q^2)&\approx& \frac{0.31}{Q^2},
\label{a17c}
\end{eqnarray}
\end{mathletters}
Eq. (\ref{a26}) is the result in Ref. \onlinecite{bos} for $x\approx 1$,
while Eq. (\ref{a25}) contains corrections for $x\ne 1$.

C) Empirical estimate, taken to be independent of $x$ (and $A$)
\cite{brad,bos,das2}:
\begin{eqnarray}
R_C(x,Q^2)\approx\frac{\delta}{Q^2}\,\,\,\,\,; 0.2 \lesssim \delta
\lesssim 0.5,
\label{a18}
\end{eqnarray}
The above, and the approximations (\ref{a22}), (\ref{a26})
for $x\approx 1$, yield $R\propto 1/Q^2$, while A) and  B) for  $x\ne 1$
prescribe $x$ dependence. It is likely that the range of extracted
$\delta$ values in (\ref{a27}) actually hides some $x$-dependence.

Were it  not for the listed uncertainties  in $\kappa^{comp}$,
the expressions (\ref{a12}), (\ref{a13}) or (\ref{a19}) would provide a
standard, against which one could test the approximate $R$ ratios A)-C).
In such a comparison, one
occasionally finds substantial differences which clearly reflect on the
extracted $F_2^A\,\,$ \cite{brad,arr1} (See Ref. \onlinecite{rt5} for details).

\section{Moments of structure functions and their relation.}

Our last  topic  regards moments of various SF
\begin{eqnarray}
{\cal M}_k^A(m;Q^2)&=&\int_0^A dx x^m F_k^A(x,Q^2)
\nonumber\\
{\cal M}_k^N(m;Q^2)&=&\int_0^1 dx x^m F_k^N(x,Q^2)
\nonumber\\
\mu^A(m;Q^2)&=&\int_0^A dx x^m f^{PN}(x,Q^2)
\label{a19}
\end{eqnarray}
The moments ${\cal M}_k^N$  measure higher twist corrections in  SF of
nucleons \cite{pen} and the same holds for their nuclear
counterparts, had those been calculated in QCD. Our interest here lies in their
sensitivity  for large $x$ and consequently the
trust in calculated $F^A_k$ for that range.
One derives from (\ref{a2}) $\,$ \cite{foot2}
\begin{mathletters}
\label{a20}
\begin{eqnarray}
F_k^A(0,Q^2)&=&\mu^A(-2+k;Q^2)F_k^N(0,Q^2)
\label{20a}\\
{\cal M}_k^A(m,Q^2)&=&\mu^A(m-1+k;Q^2){\cal M}_k^N(m;Q^2)
\label{a20b}\\
\mu^A(m+1;Q^2)&=&\frac {{\cal M}_1^A(m+1;Q^2)}{{\cal M}_1^N(m+1;Q^2)}
=\frac {{\cal M}_2^A(m;Q^2)}{{\cal M}_2^N(m;Q^2)}
\label{a20c}
\end{eqnarray}
\end{mathletters}
Eq. (\ref{a20c}) for $m$=-1
\begin{eqnarray}
\mu^A(0,Q^2)=\int_0^A dx f^{PN}(x,Q^2)=\int_0^A dx f^{as}(x)=1,
\label{a21}
\end{eqnarray}
expresses  unitarity, whereas the relations (\ref{a29})-(\ref{a31})
for finite
$Q^2$ rest  on the  representation (\ref{a2}) and  embody effects  of the
binding medium on moments of $F_k^N$ through $\mu(n,Q^2)$.  For instance,
the deviation  of $\mu^A(2,Q^2)$  from 1 measures  the difference  of the
momentum fraction  at given $Q^2$ of a quark  in a nucleus and  in
the nucleon.

We  have calculated  the  lowest  moments ${\cal M}_k$ and their ratios
$\mu$ from  computed
$F_k^A, f^{PN}$ and parametrized  $F_k^N$.
With  inaccuracies in $F_k^A$ growing with $x$ (say for $x\gtrsim  1.2$)
one  expects moments to get less trustworthy for increasing order $m$.
Yet we  found consistency between   ratios of moments of $F^A, F^N$, Eq.
(\ref{a30}), up to $m=4$ and and their corresponding ratios $\mu$,
Eq. (\ref{a31}). Fig. 3 shows  reasonable agreement  with available Fe
data.  We note in  particular the  rendition of  the observed
$Q^2$-dependence. Cothran et al, also used
the generalized convolution with $f^{PN}$ in (\ref{a2}) in the
$Q^2$-independent  PWIA, and naturally find the same results for
$\mu(m)$. Estimates for off-shell nucleons  produced far too
small moment ratios with, moreover  incorrect $Q^2$ behaviour \cite{coth}.

Modifications  of the nucleon SF in a binding medium as expressed by
(\ref{a29})-(\ref{a31}),
are reminiscent of proposals to attribute discrepancies
between data and computed results for relatively low-$q$, longitudinal
responses  $S_L$ as well as for the integral  of the latter, the
Coulomb sumrule \cite{mez,coh}. Those have  occasionally   been  ascribed
to  the influence of  the binding medium  on the size of  a nucleon, i.e.
on the second  moment  of the  $static$  charge  density (see
\onlinecite{jour} for possible conventional explanations).  One notes
that Eqs.  (\ref{a2}) and (\ref{a29})-(\ref{a31}) relate  to (moments of)
dynamical SF and  not to static form factors, from  which one obtains the
rms radii of nucleons.

In summary, we applied model  calculations of nuclear structure functions
$F_k^A(x,Q^2)$ at  relatively high $Q^2$  to a number of  observables, as
are  inclusive  cross sections,  $R$-ratios  and  moments of  SF.   Those
observables  for   variable  $Q^2$  are  sensitive   to  quite  different
$x$-ranges, not  all of which  can be computed with  comparable accuracy.
It is gratifying to note good  agreement between predictions and data for
these observables.  Of  course, that agreement is no proof  for the basic
conjecture (\ref{a2}) but it certainly supports it.

The fact that the computations avoid any element of QCD, while manifestly
being  related to  those, makes  one  wonder whether  the above  relation
results  from an  effective theory.   We have  as yet  no answer  to this
intriguing question.

\begin{center}
{\bf Table I}
\vskip 1cm
\begin{tabular}{|c|c||c|c||c|c|}
\hline
$\langle y_G\rangle$ (GeV)   &   $\theta$   & $x$  & $Q^2$ (GeV$^2$)  &
$\xi^{C,Fe}$ &$\xi^{Fe,Au}$ \\
\hline
\hline
     &  23 & 2.30 &     2.26    &  0.81 & 1.03   \\
-0.4 &  30 & 1.95 &     3.38    &  0.70 & 0.84   \\
     &  45 & 1.67 &     5.46    &  0.97 &  -     \\
\hline
     &  15 & 2.49 &     1.05    &  0.82 & 1.00   \\
-0.2 &  30 & 1.37 &     3.09    &  0.98 & 1.19   \\
     &  55 & 1.30 &     5.78    &  0.87 & 1.24   \\
\hline
     &  15 & 1.02 &     0.97    &  1.18 & 1.05    \\
 0.0 &  30 & 1.01 &     2.79    &  1.04 & 1.16    \\
     &  74 & 1.01 &     5.77    &  1.28 & 0.84    \\
\hline
     &  15 & 0.65 &     0.91    &  0.97 & 1.02    \\
 0.2 &  30 & 0.72 &     2.43    &  1.00 & 1.10    \\
     &  74 & 0.74 &     4.54    &  1.08 &   -     \\
\hline
 0.4 &  15 & 0.43 &     0.83    &  1.00 & 1.03    \\
\hline
\end{tabular}
\end{center}
\vskip 1cm
Selection of  cross section ratios $\xi^{A_1,A_2}$,
Eq. (\ref{a4}).  For each selected, narrowly-binned $\langle y_G\rangle$,
available data of the ratios are given for
the smallest, some medium and largest ($x, Q^2$) in the data sets.

{Figure captions}

{Fig. 1}.
Data \cite{arr} and predictions \cite{rt2,rt3} for the CEBAF 89-008
experiment.

{Fig. 2}.
The NE part of the GRS-type  scaling function
$\xi^{Fe,N(NE)}(y_G) \le 0,Q^2)$ (\ref{a7a})
as function of $Q^2$. Dots connect extracted values, drawn
lines data, with calculated NI parts removed.

{Fig. 3}.
Second, third and fourth moments $\mu(m,Q^2)$, Eq. (\ref{a20c}).


\bigskip


\begin{references}


\bibitem{arr}
J.R. Arrington et al, Phys. Rev. Lett. 82, 2056 (1999).

\bibitem{day}
D.B. Day $et\, al$, Phys. Rev. C 48, 1849 (1993).

\bibitem{brad}
B.W. Fillipone, $et\,al$, Phys. Rev. C 45, 1582 (1992).

\bibitem{arr1}
J. Arrington $et\, al$, Phys Rev. C 53, 2248 (1996).

\bibitem{gr}
S.A. Gurvitz and A.S. Rinat, TR-PR-93-77/
WIS-93/97/Oct-PH; Progress in Nuclear and Particle Physics,

\bibitem{aku}
S.V. Akulinitchev et al, Phys. Lett. B 158, 485 (1985); Phys. Rev.
Lett. 59, 2239 (1985); Phys. Rev. C 33, 1551, (1986).

\bibitem{bod}
A. Bodek and J. Ritchie, Phys. Rev. D23, 1070 (1981); P. Amadrauz
$et\,al$, Phys. Lett. B 295, 159 (1992); M. Arneodo $et\, al\,,ibid$
B 364, 107 (1995).

\bibitem{rt1}
A.S. Rinat and M.F. Taragin, Nucl. Phys. A 598, 349 (1996).

\bibitem{rt2}
A.S. Rinat and M.F. Taragin, Nucl. Phys. A 620, 417 (1997);
Erratum: Nucl. Phys. A 623, 773 (1997).

\bibitem{grs}
H.A. Gersch, L.J. Rodriguez and Phil N. Smith,
Phys. Rev. A 5, 1547 (1973); H.A. Gersch and L.J. Rodriguez,
Phys. Rev. A 8, 905 (1973).

\bibitem{rt3}
A.S. Rinat and M.F. Taragin, Phys. Rev. C 60, to be published.

\bibitem{om}
O. Benhar, A. Fabricioni, S. Fantoni, G.A. Miller,
V.R. Pandharipande and I. Sick, Phys. Rev. C 44, 2328 (1991);
Phys. Lett. B 359, 8 (1995).

\bibitem{ciof1}
C. Ciofi degli Atti, E. Pace and G. Salm$\grave e$, Phys. Rev. C 43,
1155 (1991).

\bibitem{ciof2}
C. Ciofi degli Atti, D.B. Day and S. Liuti, Phys. Rev.
C 46, 1045 (1994).

\bibitem{oset}
P. Fernandez de Cordoba, E. Marco, H. Mutter, E. Oset and A. Faessler,
Nucl. Phys. A 611, 514 (1996).

\bibitem{ciof3}
C. Ciofi degli Atti and S. Simula, Phys. Lett.
B 325, 276 (1994).

\bibitem{rj}
A.S. Rinat and B.K. Jennings, Phys. Rev. C 59, June issue.

\bibitem{sag}
S.A. Gurvitz, Phys. Rev. C 42, 2653 (1990).

\bibitem{day1}
D.B. Day, J.S. McCarthy, T.W. Donnely and I. Sick, Ann. Rev. of Nucl. and
Particle Physics,  40, 357 (1990).

\bibitem{rt4}
A.S. Rinat and M.F. Taragin, Phys. Lett. B, to be published.

\bibitem{atti}
C. Ciofi degli Atti, D. Faralli and G.B. West, Elba Workshop on Electron-
Nucleus scattering, EIPC, June 1998; C.Ciofi degli Atti and G.B. West,
 nuc-th 9905013.

\bibitem{arneo}
M. Arneodo, Phys. Rep  240, 301 (1994).

\bibitem{das1}
S. Dasu $et\,al$, Phys. Rev. Lett. 60, 2591 (1988).

\bibitem{berg}
J.P. Berge $et\,al$, Zeitschr. f. Physik C 49, 187 (1991).

\bibitem{bos}
P.E. Bosted $et\,al$, Phys. Rev. C 46, 2505 (1992).

\bibitem{das2}
S. Dasu $et\,al$, Phys. Rev. Lett. 61, 1161 (1988); Phys. Rev. D 49, 5641
(1994).

\bibitem{rt5}
A.S. Rinat and M.F. Taragin, submitted to Phys. Rev. C.

\bibitem{pen}
M.R. Pennington and G.G. Ross, Nucl. Phys. B  179, 324 (181);
L.F. Abbott and R.M. Barnett, Ann. of Phys.  125, 276 (1980).

\bibitem{foot2}
We disregard  here the so-called
flux  factor,  holding  that  $f(z)\to  f(z)/z$,  or  equivalently,
$\mu(m+1)\to \mu(m)$.   Numerical consequences  are anyhow  minute, since
$\mu(m)$ changes only gently with  $m$.

\bibitem{coth}
C.D.F. Cothran, D.B. Day and S. Liutti, Phys. Lett. B  429, 46 (1998).

\bibitem{mez}
Z.E. Meziani, Nucl. Phys. A 466, 113c (1985).

\bibitem{coh}
T.D. Cohen, J.W. van Orden and P. Picklesimer, Phys. Rev. Lett.  57, 1267
(1987).

\bibitem{jour}
J. Jourdan, Nuc. Phys. A 603, 117 (1996).
\end{references}
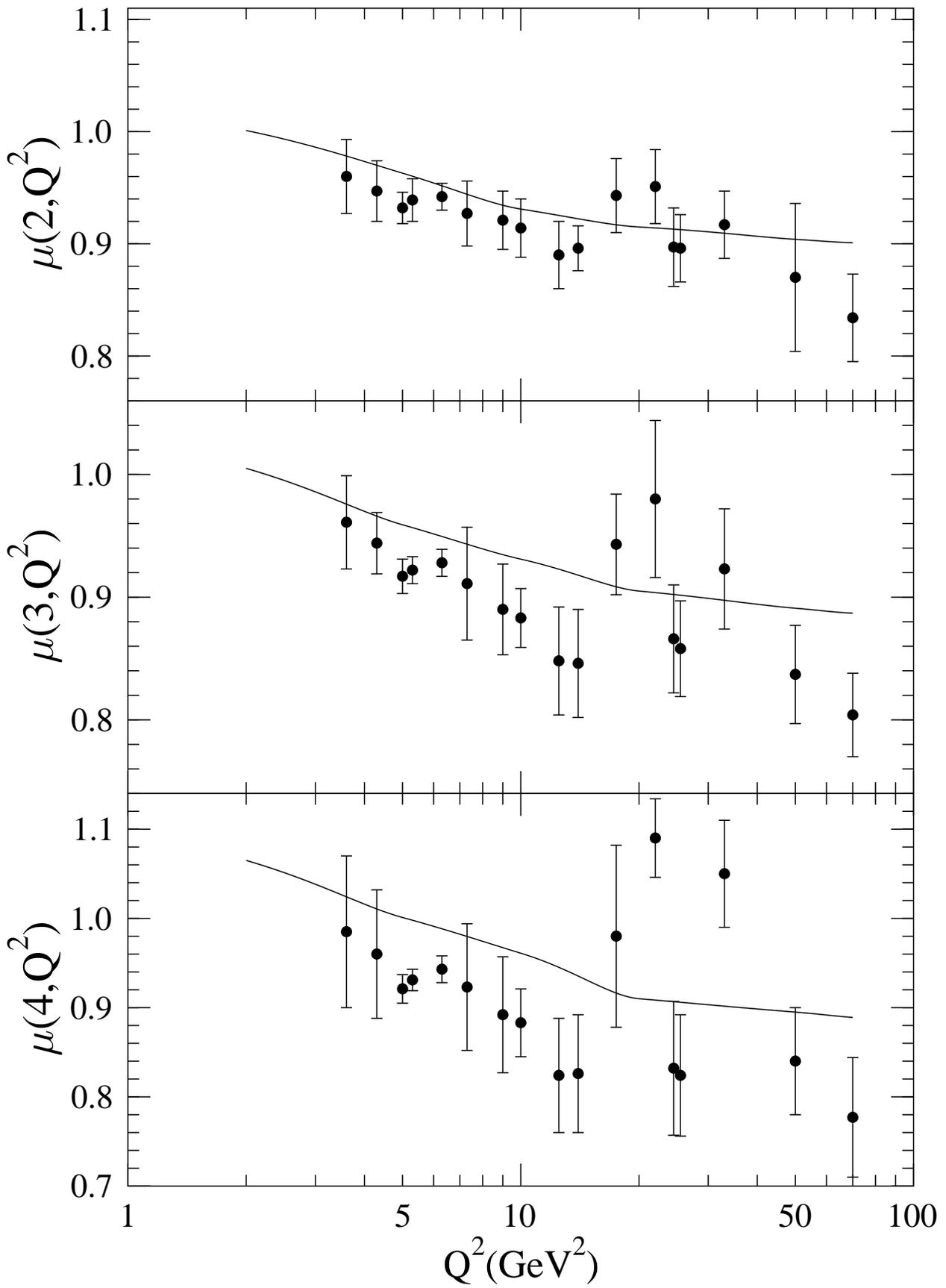
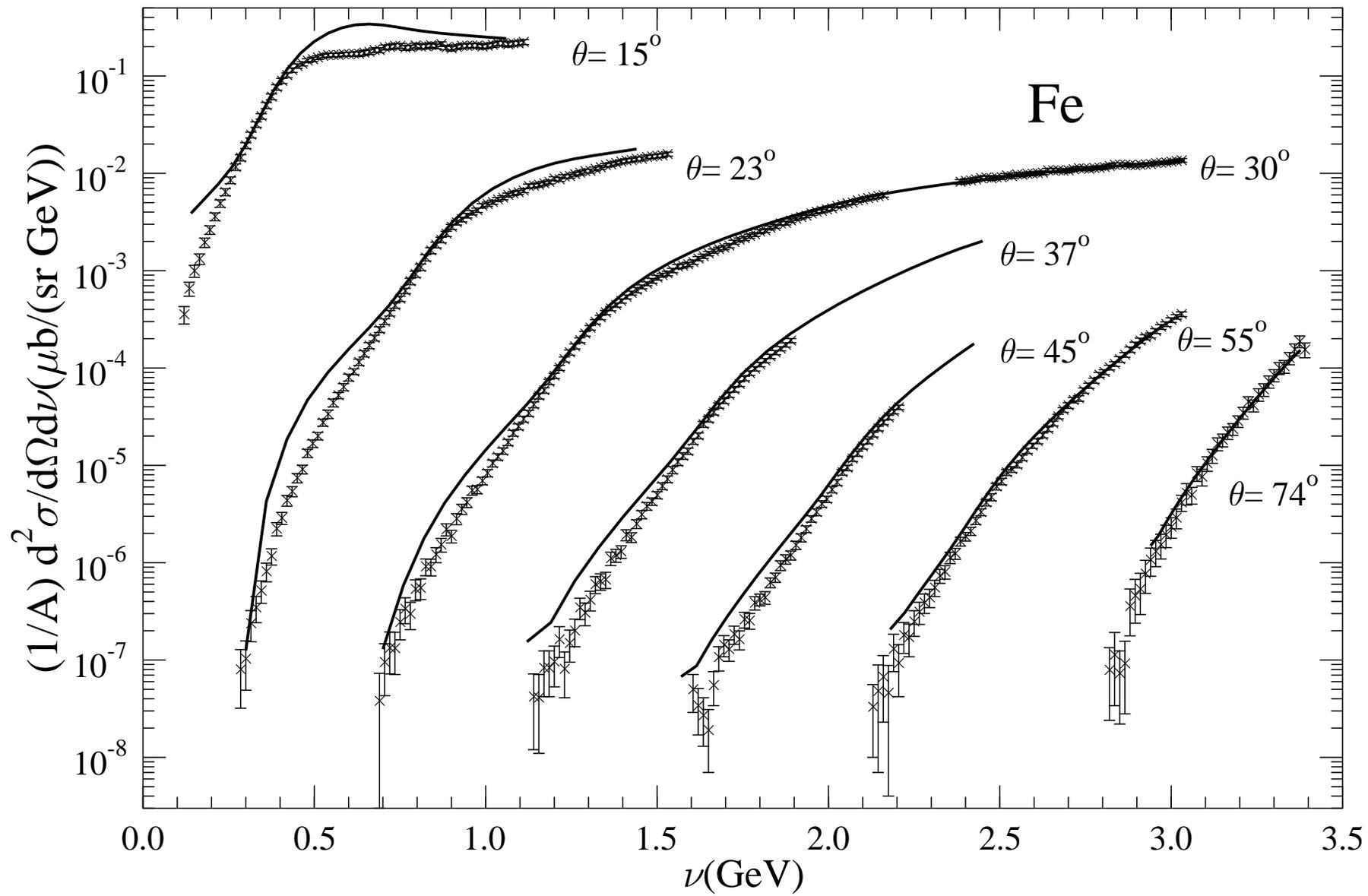
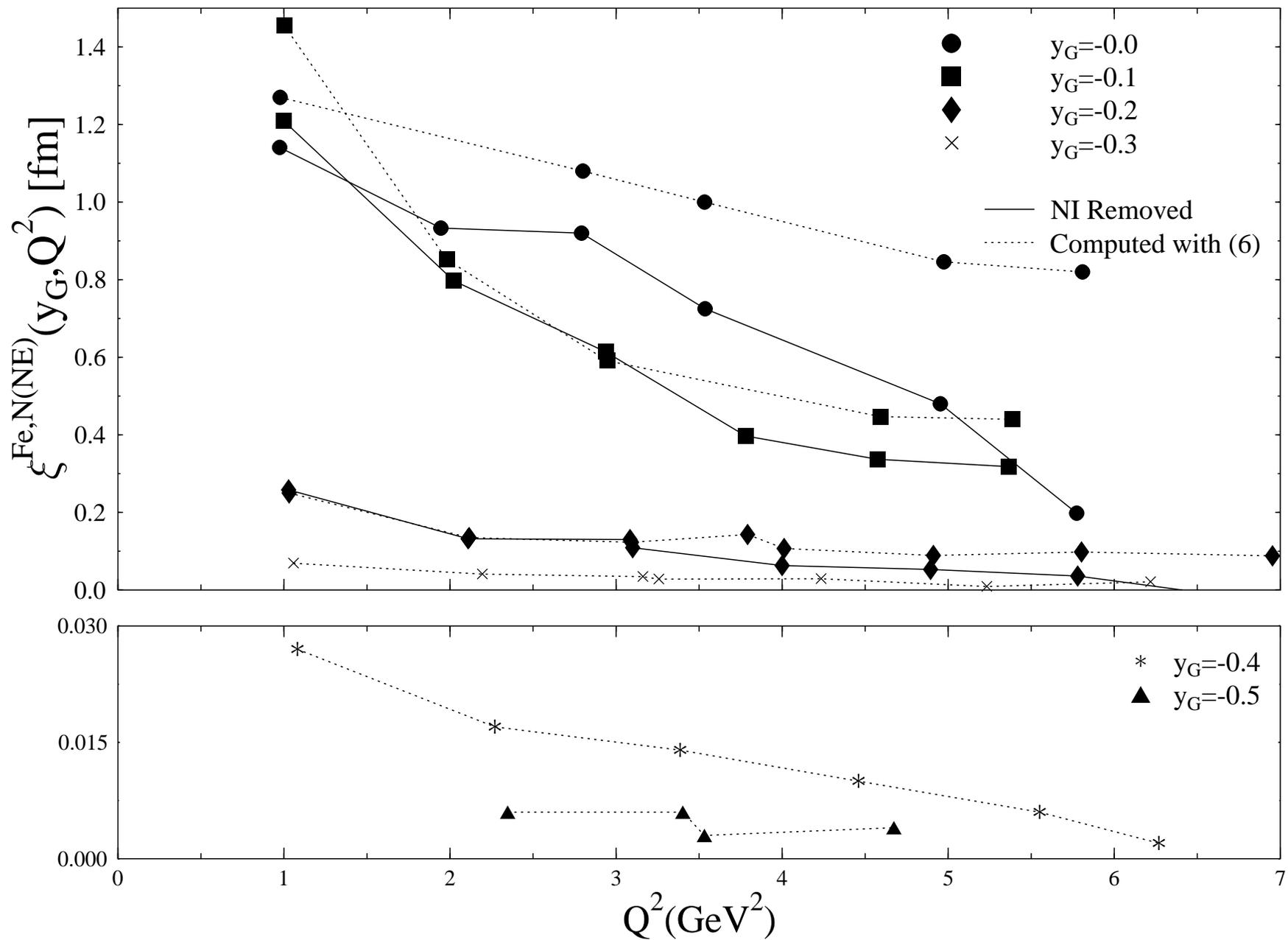
\end{document}